\documentclass[conference]{IEEEtran}
\IEEEoverridecommandlockouts
\usepackage{cite}
\usepackage{amsmath,amssymb,amsfonts}
\usepackage{algorithmic}
\usepackage{graphicx}
\usepackage{textcomp}
\usepackage{xcolor}
\def\BibTeX{{\rm B\kern-.05em{\sc i\kern-.025em b}\kern-.08em
    T\kern-.1667em\lower.7ex\hbox{E}\kern-.125emX}}
    
\makeatletter
\newcommand{\linebreakand}{%
  \end{@IEEEauthorhalign}
  \hfill\mbox{}\par
  \mbox{}\hfill\begin{@IEEEauthorhalign}
}
\makeatother

\begin{document}

\title{Batching Circuits to Reduce Compilation in Quantum Control Hardware \\
\thanks{US Department of Energy, Office of Science, Office of Advanced Scientific Computing Research Quantum Testbed Program}
}

\author{\IEEEauthorblockN{Ashlyn D. Burch}
\IEEEauthorblockA{\textit{Sandia National Laboratories}\\
Albuquerque, NM, USA\\
adburch@sandia.gov} 
\and
\IEEEauthorblockN{Daniel S. Lobser}
\IEEEauthorblockA{\textit{Sandia National Laboratories}\\
Albuquerque, NM, USA \\
dlobser@sandia.gov} 
\and
\IEEEauthorblockN{Christopher G. Yale}
\IEEEauthorblockA{\textit{Sandia National Laboratories}\\
Albuquerque, NM, USA \\
cgyale@sandia.gov}
\linebreakand
\IEEEauthorblockN{Jay W. Van Der Wall}
\IEEEauthorblockA{\textit{Sandia National Laboratories}\\
Albuquerque, NM, USA \\
}
\and
\IEEEauthorblockN{Oliver G. Maupin}
\IEEEauthorblockA{\textit{Dept. of Physics and Astronomy at} \\
\textit{Tufts University}\\
Medford, MA, USA \\
}
\and
\IEEEauthorblockN{Joshua D. Goldberg}
\IEEEauthorblockA{\textit{Sandia National Laboratories}\\
Albuquerque, NM, USA \\
}
\linebreakand
\IEEEauthorblockN{Matthew N. H. Chow}
\IEEEauthorblockA{\textit{Sandia National Laboratories}\\
\textit{Center for Quantum Information and Control}\\
\textit{Dept. of Physics and Astronomy at}\\
\textit{University of New Mexico}\\
Albuquerque, NM, USA \\
}
\and
\IEEEauthorblockN{Melissa C. Revelle}
\IEEEauthorblockA{\textit{Sandia National Laboratories}\\
Albuquerque, NM, USA \\
}
\and
\IEEEauthorblockN{Susan M. Clark}
\IEEEauthorblockA{\textit{Sandia National Laboratories}\\
Albuquerque, NM, USA \\
sclark@sandia.gov}
}


\maketitle

\begin{abstract}
At Sandia National Laboratories, QSCOUT (the Quantum Scientific Computing Open User Testbed) is an ion-trap based quantum computer built for the purpose of allowing users low-level access to quantum hardware. Commands are executed on the hardware using Jaqal (Just Another Quantum Assembly Language), a programming language designed in-house to support the unique capabilities of QSCOUT. In this work, we describe a batching implementation of our custom software that speeds the experimental run-time through the reduction of communication and upload times.  Reducing the code upload time during experimental runs improves system performance by mitigating the effects of drift. We demonstrate this implementation through a set of quantum chemistry experiments using a variational quantum eigensolver (VQE). While developed specifically for this testbed, this idea finds application across many similar experimental platforms that seek greater hardware control or reduced overhead. 
\end{abstract}

\begin{IEEEkeywords}
quantum computation, ion traps, quantum control hardware
\end{IEEEkeywords}

\section{Introduction}

Noisy Intermediate Scale Quantum (NISQ) computers are currently the dominant type of quantum hardware available to the scientific community.  There are various commercially available hardware systems~\cite{ibm, rigetti, ionq, quantinuum} as well as the Department of Energy Quantum Testbeds~\cite{qscout, aqt} that researchers can use to help answer questions in quantum computing,  quantum chemistry, and quantum simulation, among others.  An important piece of quantum system performance is the classical control hardware and software, including the subtle interactions between these classical systems.  In this work, we discuss a specific batching modification to our software, allowing us to upload multiple sets of circuits at once and run through the sequence of circuits on the hardware rather than compiling and running a single circuit at a time. This was necessary for two reasons: 1) spatial drifts called for frequent recalibration, thus making it unreasonable to execute code with a longer upload time, and 2) on-chip memory buffers created limitations for the number of circuits that could be run.  We demonstrate the batching implementation through execution of variational quantum eigensolver (VQE) simulations to find the ground state energy of the diatomic molecule HeH$^+$. This implementation results in an overall improved performance of our quantum machine by allowing it to run more circuits before the system requires recalibration. 

In general, one of the greatest challenges with NISQ devices is the impact of uncontrolled noise on qubit operations. This concern increases with the number of qubits in the system. Even small platforms can be affected by the negative impacts of noise and drift, making them good candidates for improving speed performance while the number of qubits remains lower. 

The system used in this study is the Quantum Scientific Computing Open User Testbed (QSCOUT)~\cite{b1}, which is based on trapped ions and operated at Sandia National Laboratories.   QSCOUT is specifically designed to provide greater access to a quantum circuit through parameterized one- and two-qubit gates as well as lower level pulse control if desired. In order to offer these capabilities, we developed Just Another Quantum Assembly Language (Jaqal) ~\cite{b2, b3} for circuit construction. Jaqal was designed such that users have the ability to customize their code based on the native built-in ion-trap quantum gates as well as have the functionality to specify those gates in serial or parallel sequences. Current capabilities on QSCOUT allow for control of up to 5 qubits, with plans of scaling to 32.  Platforms like QSCOUT are the ideal system for in-depth investigation into the nature of low-level noise and errors and have been used to determine new directions to increase performance in the presence of these noise levels \cite{richerme_2022, majumder_2022}.

\section{Software - Jaqal}

\subsection{Jaqal}

As implemented on QSCOUT, the Jaqal programming language provides the following native operations: preparation of all qubits in the z-basis, parameterized single-qubit rotation gates about any axis on the equatorial plane of the Bloch sphere, single-qubit virtual $R_Z$ gates, arbitrary-angle M\o lmer-S\o rensen (MS)~\cite{MS} gates between any two pairs of ions, and measurement of all qubits in the z-basis. These gates are not fundamental to the Jaqal language but are defined in a gate pulse file (GPF) in JaqalPaw~\cite{jaqalpaw}, the pulse-level counterpart of Jaqal. These GPFs can be thought to build a ``standard library" for Jaqal quantum operations. Beyond these native gates, users have the ability to customize their code by defining macros, which creates new composite gates from a specified sequence of native gates. This allows for gates not directly implementable in ion trap hardware such as a Hadamard, controlled-NOT, or ``\emph{ZZ}" MS gate. Each gate is executed as a pulse sequence defined in a GPF. The gate definitions are sometimes updated to improve gate calibration and performance or to accommodate new types of native gates.

In addition to macros, Jaqal circuits make use of let statements to define integer or floating point values. Once set, these let parameters can be used throughout the code as identifiers. Besides making the source code easier to read, let parameters can be efficiently changed by the QSCOUT control software without requiring a full recompilation of the Jaqal source file.

\subsection{JaqalPaq}

While Jaqal files can be written directly by a user, they do not allow for higher-level classical control or metaprogramming. For example, there are no if statements or for loops native to Jaqal, either for classical variables or the result of quantum measurements. Jaqal is therefore an intentionally simple language. To express complex ideas with such a simple medium we created JaqalPaq~\cite{jaqalpaq}. JaqalPaq is a Python software package containing several sub-packages including a compiler, code generator, emulator, and programmatic circuit creation. The latter capability can be used to programmatically produce a Jaqal file, or many similar Jaqal files, using the full expressiveness of Python.

\subsection{Jaqal Application Framework}

\begin{figure*}
\centering{\includegraphics[width=190mm]{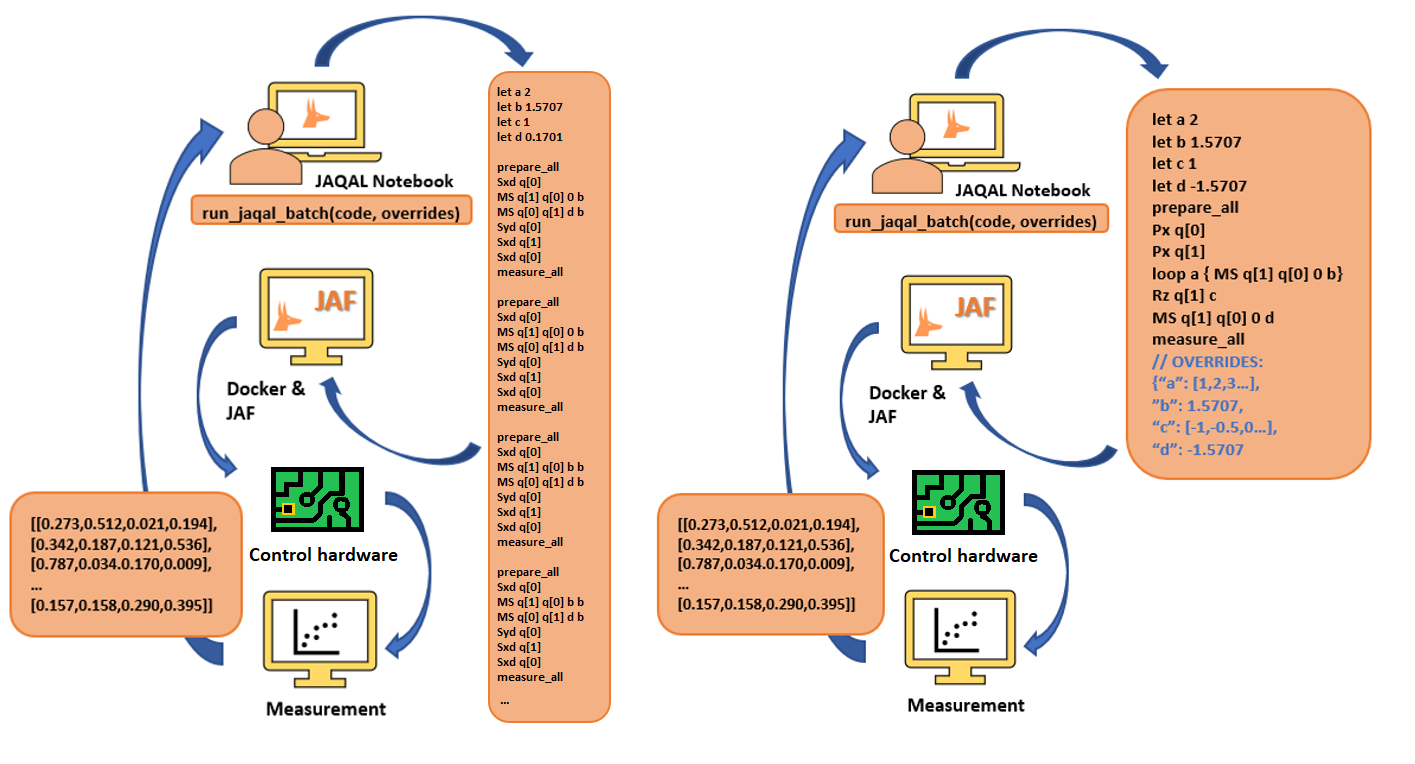}}
\caption{Batching in JAF allows for multiple Jaqal files to be run on the QSCOUT control hardware without calling back to the Jaqal notebook for each variation in the code. In one configuration, displayed on the left, this was accomplished by batching through indexing, where indices were coupled to parameter overrides. In this configuration, the Python generator reduces the amount of compilation overhead and allows for more circuits to be run at a time, as opposed to previous implementations in JAF which fed each set of \emph{prepare\_all} and \emph{measure\_all} circuits individually. The code is written in the Jaqal Notebook and is run using the \emph{run\_jaqal\_batch} command. This code is then uploaded to JAF, which communicates with the hardware to perform the measurement. The final results are sent back to the main notebook to compare to emulated results. A similar batching method, displayed on the right, was accomplished by creating a let parameter dictionary that could act as a set of overrides for each parameter. As before, the code is written in the Jaqal Notebook and is run using the \emph{run\_jaqal\_batch} command, calling the corresponding code with its list of dictionary overrides.}
\label{figbatch}
\end{figure*}

The Jaqal Application Framework (JAF) is a software framework that was created to facilitate executing Jaqal circuits on the QSCOUT hardware. It comprises a network service running on a Docker container, an adapter layer in the hardware's control software, and a GUI used to select the application to run. This architecture abstracts the different components from each other. From a maintenance perspective, this prevents failures in either JAF or the QSCOUT system from affecting the other. It also allows new capabilities to be built into JAF with minimal changes to other systems.

An application in JAF can be either a plain Jaqal text file, a Python program, or a Jupyter notebook. If the application is a Python program or Jupyter notebook, it must use JaqalPaq to programmatically create and request execution of a Jaqal circuit. JaqalPaq and JAF work together to abstract the execution back end from the application producing the Jaqal circuit. This means that an application will work with either the JaqalPaq emulator or the QSCOUT hardware without modification. Additionally, the QSCOUT hardware is generally unaware of the specific type of application being run; it is given a Jaqal circuit with some metadata to run and returns the averaged result of each measurement call.

When running a circuit, JAF returns result by subcircuit. A subcircuit is defined as zero or more gates bounded by a preparation and measurement call, \emph{prepare\_all} gate and \emph{measure\_all} gate, respectively. A circuit consists of one or more subcircuits. These subcircuits are either implicit, denoted by the presence of the \emph{prepare\_all} and \emph{measure\_all} gates, or explicit with the subcircuit Jaqal keyword.

\subsection{Batching in Jaqal}


In initial implementations, in any routine requiring multiple circuits to be run, JAF would feed each individual circuit to the experiment and wait for a response before sending the next. This resulted in a slower execution time due to the overhead associated with each pulse-level compilation and multiple network transfers. By default, pulse-level compilation of Jaqal code fetches the latest experimental calibration data and regenerates all of the binary data needed for executing gates on the QSCOUT control hardware.  We call the binary data ``bytecode" in analogy with interpreted programming languages. 

 When multiple circuits have a large amount of overlap in gate data and when calibration parameters are not expected to change (e.g. all circuits utilize the same subset of gates but in different configurations), significant speedup can be achieved by compiling some or all of the circuits upfront. This process is demonstrated in Fig.~1, showing how multiple Jaqal files can be fed into and run on the software without having to connect back to the Jaqal notebook each time there is a variation applied to the circuit. In this first case, the compressed bytecode representation for all circuits can be identical to, or perhaps marginally larger than, the bytecode for a single circuit, depending on the amount of inter-circuit variation in gate calls. Thus, multiple circuits can be calculated once and executed by uploading only minimal sets of bytecode. The subcircuits from the compressed representation are then used to stream out RF pulses from the control hardware. 

Another use case exists where circuits are identical but need to be run using different values for calibration parameters or Jaqal let parameters. Here, we employed a different approach in order to reduce the effective overhead of compilation. The Jaqal code is passed once per batch, and the let parameters are passed in a separate dictionary to specify their override values. The dictionary keys correspond to the parameter names, and the values are either: 1) scalar, in which case the override is applied for every run in the batch, or 2) an array, which defines multiple override values to be applied sequentially on consecutive runs. When multiple parameters are specified using arrays, their values are tied together so that each array is iterated over with a common index, and thus their lengths must match. 

The low-level bytecode is subject to potentially large changes, depending on the parameters being updated. When these parameters are known in advance, we can achieve similar speedup by wrapping the entire compilation step into a Python generator that iteratively queues up multiple output products. This offsets initial compilation wait time at the beginning of code execution with experimental run time. Using a generator also allows for arbitrarily large numbers of circuits to be run because blocking conditions during upload caused by maximal filling of on-chip buffers can throttle compilation. Rather than doing a compilation every time a new set of code is passed in, all of the compilation now happens upfront. 

Because of the differences in how batching is executed for multiple circuits through indexing and batching execution of parameter overrides are treated, we can also get compilation performance gains when combining the two. However, this multiplicative gain only applies if multiple subcircuits all use the same parameter overrides. In that case, they are executed once. If subcircuit indices are instead coupled with unique parameter overrides, or the overrides are iterated over one subcircuit index at a time, then the performance gain is comparable to a single circuit with override batching.

\section{Hardware - QSCOUT}

The QSCOUT system is a room-temperature trapped-ion quantum computer that exploits the hyperfine states of Ytterbium ($^{171}$Yb$^+$) ions. The ions are arranged in a one-dimensional linear chain, and they are fully connected through their vibrational modes. They are trapped in a Sandia-fabricated High Optical Access (HOA2.1)~\cite{maunz} surface-electrode trap, using a radio-frequency (RF) trapping pseudopotential. 
We image each individual ion into a single core from a multicore fiber array, with each core then coupled to individual PMTs for distinguishable detection of the qubit state.

Using a pulsed 355 nm laser, we excite Raman transitions to perform single- and two-qubit gates between any pair of the ions. The 355 nm light is split into two beams and applied to the ions in a counterpropagating configuration. One of the beams is the global beam, which encompasses all ions, while the other becomes a series of tightly focused individually-addressing beams. Frequency tones are applied to various sets of beams in order to generate the appropriate Raman transitions for single- and two-qubit gates. The single-channel acousto-optic modulator (AOM) used to generate the global beam and the multi-channel AOM used to generate the individually addressing beams are both controlled by the experiment hardware, generating up to two RF tones on each channel within the AOMs. 

For a single-qubit gate about an equatorial axis, we apply two tones to a single individual addressing beam, with the duration of the pulse determining the overall angle traversed. Likewise, a phase, or $R_Z$ gate is applied virtually, existing as a phase shift on all subsequent waveforms for that particular qubit. The native two-qubit gate within the trapped-ion system is the MS gate, an entangling gate which couples the ions via their vibrational modes, forming an  $e^{-i\frac{\theta}{2} \sigma_{X} \otimes \sigma_{X}}$, or ``\emph{XX}," interaction. This gate is generated via Raman transitions which are symmetrically detuned from both a red and blue motional sideband, driven by tones on the individual beams addressing the ions of interest and a third shared tone on the global beam. The maximally entangled state occurs at $\theta = \pm \pi/2$, and these are the gates present in the circuit used for this demonstration. To physically generate the negative angle, $\theta = -\pi/2$, we shift the interaction from ``\emph{XX}" to ``\emph{X,-X}" by shifting the phase of the waveforms addressing one of the two ions by $\pi$ radians.

\section{Experimental Results}

One example algorithm that takes advantage of these speedups due to batching is a variational quantum eigensolver (VQE) algorithm.  VQE is used to model the behavior of energetic characteristics of molecules and is used in quantum chemistry problems for extracting the upper bound ground state energy of a Hamiltonian~\cite{vqe}. One of the unique features of VQE is that it is a quantum and classical hybrid, utilizing a quantum processor with a classical optimizer. 

In this experiment, to demonstrate the experiment speed-up due to the reduction in compilation, we performed VQE using the constrained optimization by linear approximation (COBYLA), a classical numerical optimization method that is used to find the minimum of the Hamiltonian. The circuits for these experiments all contain the following configuration:  [$XX(\frac{\pi}{2})R_Z(\theta)XX(-\frac{\pi}{2})$], two MS gates separated by a virtual Z gate, where $\theta$ is tuned in the quantum circuit to construct the ansatz. For each evaluation of the ground state energy of HeH$^{+}$, we make estimations of nine different Pauli terms (which we refer to here as projections) that comprise the Hamiltonian through averaging the results of the experimental outcomes. The optimizer takes this calculated energy and vector of parameters and gives an updated parameter for the next iteration by making and updating a linear approximation of the objective function through repeated evaluations. The algorithm then again refines its search at higher resolutions. 

\begin{figure}[htbp]
\centerline{\includegraphics[width=95mm]{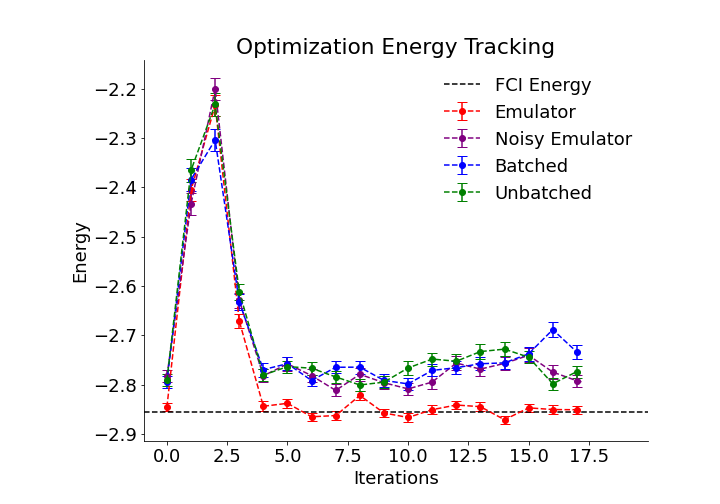}}
\caption{Results of the VQE algorithm using the COBYLA optimizer, depicting results prior to batching as well as using the new batching framework in JAF. With no batching, 17 iterations equated to a total of 162 steps (9 projections with 18 iterations each), while the new regime with batching allows all 9 projections to be stored in the let parameter override dictionary resulting in a total of 18 steps, where each batch is an optimizer step.}
\label{figenergy}
\end{figure}

The results from these measurements as well as the emulated values are plotted in Fig. 2. The outcomes were averaged over 1000 shots. We used 18 iterations to calculate the energy, although this value could have been extended even further for more optimizer calculations. As there are nine different projections in this algorithm, the experiment would use a total of 162 steps, 9 steps per optimizer iteration, each time communicating with the Jaqal Notebook to create a new Jaqal file to be run on the hardware. The average run time for taking this set of data took 29 minutes. After implementing the changes to Jaqal using batching, each batch step equated to one optimizer step as all 9 projections were now accessed through the let parameter overrides (refer to Fig.~\ref{figbatch}). This instead totaled 18 communication steps, 1 for each iteration of the optimizer. The communication time between each new Jaqal file and JAF takes about two seconds, and this decrease in the number of steps decreased the run-time of the experiment, only taking on average 19.5 minutes in the new configuration.

The noisy emulator is calculated using a simulated error model for the gates that is meant to mimic errors in experimental control. These parameters include the power, frequency, phase, and timing error of the gate laser pulses as measured by their distances from the ideal gate, as well as the Gaussian width of each of these errors. We modeled the behavior of these gates in simulation using process matrices adjusted according to the respective parameters. This model does not incorporate the effects due to regular detuning fluctuations or time-averaged drifts, both of which are problems which occur in longer experiments. Both sets of results show a good match with the noisy emulator, showing that our batching technique produces comparable results in a shorter amount of time. 



\section{Discussion}

In the current version of QSCOUT, a dominant error is drift due to fluctuations in RF power that create the pseudopotential and define the secular frequencies. For the experimental results presented in Fig. 2, there is little variation between the batched and unbatched VQE data. On the days these measurements were taken, the initial and final fidelities were comparable and drift was minimal. However, due to a recent shorting event on the trap that greatly affected RF stability, QSCOUT has been exposed to more drift than was reported in \cite{b1}. Fig. 3 is an example of calibration data that was recorded on a day measurements were greatly affected by RF power fluctuations. Here, we plot the red sideband vertical mode drift as a function of time. This shows a detuning drift of over tens of kHz every fifteen minutes. A drift of 3 kHz detuning on this system equates to a 2 $\pm$ 1.6$\%$ reduction in MS gate fidelity, so more intermediate recalibrations were necessary to help mitigate these effects. In longer circuit sets, these more frequent recalibrations became highly inefficient; the reduction in communication time through batching significantly improved the quality of the results and removed the necessity for intermediate recalibration. 


\begin{figure}[htbp]
\centerline{\includegraphics[width=80mm]{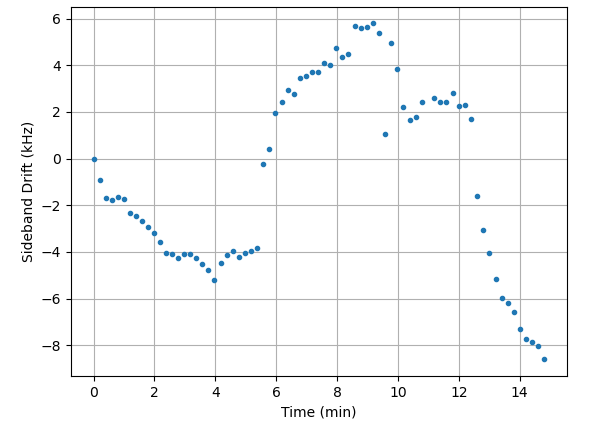}}
\caption{Example of sideband drift seen on a day measurements were greatly affected by RF power instability, showing detuning drifts of up to 15 kHz in 15 minutes. In such cases, frequent recalibrations were necessary to prevent this drift from having an effect on algorithm performance.}
\label{figenergy}
\end{figure}

Spatial drifts of the ions off of the individual addressing beams on the hour timescale can also be a source of error, affecting the performance of both single and two qubit gates. These spatial drifts lead to under- and over-rotations that also have an effect on gate fidelity. Especially for running longer circuits, the speed-up of experimental run-time that was exhibited via batching Jaqal circuits was helpful for the quality of data that was acquired. 

Here, we discussed two key batching techniques: batching quantum circuits through a dictionary of parameter overrides and batching the code through indexing. We demonstrated the first implementation of let parameter overrides while performing VQE, speeding up the experiment run-time. This new batching technique has also proved useful in other experiments where on-chip memory created limits for the number of circuits to be run, exposing the need to wrap more compilations through the use of indices. For example, in [11], we investigated randomized compiling on a VQE algorithm in the presence of noise. In randomized compiling, we average unitarily equivalent circuits with randomly selected single-qubit Pauli twirlings around the two-qubit gates in order to reduce the impact of noise. Accounting for the number of equivalent circuits to be averaged per projection, the number of projections, and other comparative circuits, we could process 90 different circuits in a single batch. In this case, each batch corresponded to a different tunable phase parameter within the ansatz. In the pre-batching regime, investigating 41 different phases would have required 3690 communication steps. With batching implemented, it instead totaled 41 steps. The communication alone for an algorithm with 3690 calls would have exceeded two hours, making such an approach infeasible on the QSCOUT system. 

For future runs of QSCOUT, we have plans to implement a new trap to help reduce the amount of drift so that longer measurements are more robust. Even so, the speed-up due to batching emphasizes the importance of careful design of control software. Having greater control over quantum hardware through software performance is not a challenge unique to this experiment, and these ideas may find application in other similar types of platforms. Even for the same circuit configuration, as demonstrated here, minute details of software implementation on quantum computers can have a drastic impact on runtime and execution performance.

\section*{Acknowledgment}

We want to acknowledge the significant contributions of Andrew Landahl, Kenneth Rudinger, Antonio Russo, and Benjamin Morrison for their work in developing and constantly improving the Jaqal programming language, making this work feasible. 

This work was supported by the U.S. Department of Energy, Office of Science, Office of Advanced Scientific Computing Research Quantum Testbed Program.
Sandia National Laboratories is a multimission laboratory managed and operated by National Technology and Engineering Solutions of Sandia, LLC, a wholly owned subsidiary of Honeywell International Inc., for the U.S. Department of Energy’s National Nuclear Security Administration under Contract DE-NA0003525. This article describes objective technical results and analysis. Any subjective views or opinions that might be expressed in the article do not necessarily represent the views of the U.S. Department of Energy or the U. S. Government. SANDXXXXXXXXXX

\end{document}